\documentclass[prb,aps,showpacs,superscriptaddress,floatfix,preprint]{revtex4}  

\usepackage{graphicx}
\usepackage{dcolumn}
\usepackage{bm}
\usepackage{amssymb,amsmath}
\usepackage[usenames]{color}
\usepackage{epsfig}

\begin{document}
\title{Kondo lattice model at half-filling}
\author{R. Nourafkan }
\affiliation{Department of Physics, Sharif University of Technology,
P.O.Box: 11155-9161, Tehran, Iran.}
\author{N. Nafari }
\affiliation{Institute for Studies in Theoretical Physics and
Mathematics, P.O.Box: 19395-5531, Tehran, Iran.}
\date{\today}
\begin{abstract}
The single- and two-channel Kondo lattice model consisting of
localized spins interacting antiferromagnetically with the itinerent
electrons, are studied using dynamical mean field theory. As an
impurity solver for the effective single impurity Anderson model we
used the exact diagonalization (ED) method. Using ED allowed us to
perform calculations for low temperatures and couplings of arbitrary
large strength. Our results for the single-channel case confirm and
extend the recent investigations. In the two-channel case we find a
symmetry breaking phase transition with increasing coupling
strength.
\end{abstract}
\maketitle
\section{Introduction}
The Kondo lattice model (KLM)\cite{Tsu}, in its single- and
two-channel forms, is one of the fundamental microscopic models for
the description of heavy fermion materials
\cite{Lee,Aeppli,Stewart,Lohneysen}. This model consists of
itinerant conduction electrons coupled to localized spins sitting on
the sites of a crystal lattice. The coupling is represented as an
on-site exchange interaction between this spin and the conduction
electron spin density. This rather simple model gives rise to
complex many body physics whose detailed understanding requires
further investigation. The nature of the ground state of this model
results from the interplay between magnetic
Ruderman-Kittel-Kasuya-Yosida (RKKY) interaction \cite{RKKY} among
the localized spins and the Kondo effect screening of these spins.
The polarization cloud of conduction electrons produced by a local
moment may be felt by another local moment. This provides the
mechanism for the RKKY interaction. On the other hand, the same
polarization cloud can also form a singlet bound state with the
local moment, when coupling is strong \cite{Hewson}. The RKKY
interaction leads to a long-range ordered antiferromagnetic phase in
two and three dimensions and the Kondo effect screening leads to
a short-range spin correlations due to the formation of coherent
Kondo spin singlets. There is a quantum phase transition between the
two limiting phases upon changing the parameters of the model
\cite{Doniach}. It is generally believed that the half-filled
single-channel Kondo lattice model (SC-KLM) exhibits a Kondo
insulator phase for large coupling strength, whereas for smaller
coupling strength a phase transition to an antiferromagnetic state
occurs \cite{Dorin}.

In contrast to the SC-KLM, which shows Fermi liquid behavior, the
two-channel Kondo lattice model (TC-KLM) shows non-Fermi liquid
behavior. By TC-KLM we means two identical species of
non-interacting electrons coupled antiferromagnetically to localized
electron spins. The materials which may display the TC-KLM are
rare-earth or actinide inter-metallic compounds such as
\(\mathrm{CeCu_{2} Si_{2}}\) and \(\mathrm{UBe_{13}}\). In these
compounds, the \(f\)-orbital of the \(\mathrm{Ce}\) and
\(\mathrm{U}\) elements remain strongly localized, essentially
retaining their atomic character. Thus, the sites containing
\(\mathrm{Ce}\) or \(\mathrm{U}\) atoms often possess a magnetic
moment obeying the first Hund's rule of maximization of the
total \(f\)-electron spins. These localized moments interact with
light conduction electron states contributed by surrounding ligands
\cite{Cox}. The single impurity case ,i.e, this model has a
non-Fermi liquid ground-state \cite{Nozieres}. Non-Fermi liquid
behavior is the result of the inability of conduction electrons to
screen the impurity spin. For weak coupling, one has a free spin
one-half object scattering electrons in both channels resulting in
the same logarithmically growing scattering as the temperature is
lowered. However, as the coupling constant grows, the impurity spin,
on the one hand, tends to form a singlet, and on the other hand, the
symmetry of the problem forbids it to favor either one of the two
channels to form that singlet. The other possibility is to make a
linear superposition of a singlet with each channel, but this leaves
the unbound spin of the spectator channel carrying a two-fold spin
degeneracy. In this case, as it turns out, it would behave like a
new spin-half impurity which in turn wants to undergo another Kondo
effect. Less is known about TC-KLM. In a study using
quantum Monte Carlo (QMC), Jarrell and coworkers \cite{Jarrell} have
examined the PM phase of this model and found non-Fermi liquid
behavior at low temperatures. The existence of sign problem in their
QMC simulation limited their access to very low temperatures and
large coupling strengths.

The aim of this work is to provide a thorough analysis of the SC-KLM
and TC-KLM at half filling. We have also studied the TC-KLM at
quarter-filling. The quarter-filling case for the two-channel model
is analogous to the half-filled case for the single-channel model in
that there is one conduction electron per impurity spin leading to
complete screening at strong couplings. Our results for the
single-channel case showed the presence of the Kondo insulator. In
the two-channel case, we find a symmetry breaking phase transition
with increasing coupling strength.

The organization of the paper is the following: In Sec. II, the KLM
is introduced. In subsections III.A and III.B, results on the PM and
AFM are presented, respectively. The main portion of subsection
III.A is devoted to the calculation of self-energies, obtained at
several coupling strengths. Subsection III.B deals with magnetic
ordering and Sec. IV contains our concluding remarks.

\section{Model}
The KLM Hamiltonian is defined by
\begin{equation} \label{eq:H}
H = -t\sum_{\substack{im
\\ \sigma}}(c^{\dag}_{im\sigma}c_{i+1,m\sigma} + H.c.) +
\frac{J}{2}\sum_{\substack{im \\ \alpha \beta}}\bm{S}_{i}\cdot
(c^{\dag}_{im\alpha} \bm{\sigma}_{\alpha \beta} c_{i+1,m\beta}).
\end{equation}
Here, \(m\) is the channel index, assuming two values \((m=1,2)\)
for the two-channel systems. \(t\) is the conduction electron
hopping amplitude, taken to be the same in both bands,
\(c^{\dag}_{im\sigma}(c_{i,m\sigma})\) creates (annihilates) an
electron on lattice site \(i\), with channel index \(m\) and spin
projection \(\sigma=(\uparrow,\downarrow)\), and \(\bm{\sigma}\) is
a pseudo-vector represented by Pauli spin matrices. \(\bm{S}_{i} \)
is the spin operator of the localized \(f\)-electrons.

For solving this hamiltonian we employ the dynamical mean field
theory (DMFT)\cite{Georges} which is a powerful tool to investigate
the nonperturbative regimes of strongly correlated systems. In DMFT,
the lattice model is mapped onto an effective impurity problem
subject to a self-consistency condition which contains the needed
information about the original lattice. The method becomes exact in
the infinite coordination limit. In order to map KLM onto an
appropriate impurity model, we closely follow the treatment given in
Ref. \onlinecite{Meyer}, i.e., we introduce fermion operators
\(f_{i\sigma}\) to represent the spin operator of the localized
\(f\)-electron \((S_{i}= \frac{1}{2})\) as
\(\bm{S}_{i}=\frac{1}{2}\sum_{\alpha,\beta}f^{\dag}_{i\alpha}
\bm{\sigma}_{\alpha \beta} f_{i\beta}\), where the \(f\)-operators
satisfy the constraint
\(f^{\dag}_{i\uparrow}f_{i\uparrow}+f^{\dag}_{i\downarrow}f_{i\downarrow}=1\)
for all \(i\). 

In the treatment of the effective impurity problem,
the numerical method often employed is the quantum Monte Carlo (QMC)
simulation based on Hirsch-Fye (H-F) algorithm and to avoid the sign
problem one considers the classical spins \cite{Furukawa}. However,
the importance of a fully quantum mechanical treatment of the local
spins has been addressed by Kienert and Nolting \cite{Kienert}.
Recently, Werner and Millis \cite{Werner} have developed the
stochastic quantum Monte Carlo (SQMC) which is based on the
stochastic evaluation of diagrammatic expantion of the partition
function. Although, the SQMC is faster than H-F's and the sign
problem is less severe, but their access to low temperatures is
limited. To avoid such limitations, we have solved the effective
impurity problem using exact diagonalization (ED) algorithm
\cite{Caffarel,ED}. The ED can handle the whole interaction and
temperature regimes. Also, due to fully quantum mechanical treatment
of the local spins, we correctly treat spin-flip processes. In the
self consistency loop of DMFT we use, as it is often done, a
semi-circular density of states of bandwidth \(4t\) (Bethe lattice).
\section{Results}
In this section we present our results on SC-KLM and TC-KLM obtained
by iteratively solving the equations appearing in DMFT.
All calculations are done at \(T=0\), unless otherwise is specified.
Moreover, to obtain the paramagnetic (PM) solutions we suppress the
magnetic order by averaging over spin up and spin down in each
orbital. In this way, we have obtained both PM and antiferromagnetic
(AFM) DMFT solutions of the KLM.
\subsection{Pramagnetic Phase}
\begin{figure}[htb]
\includegraphics[width=10cm,clip]{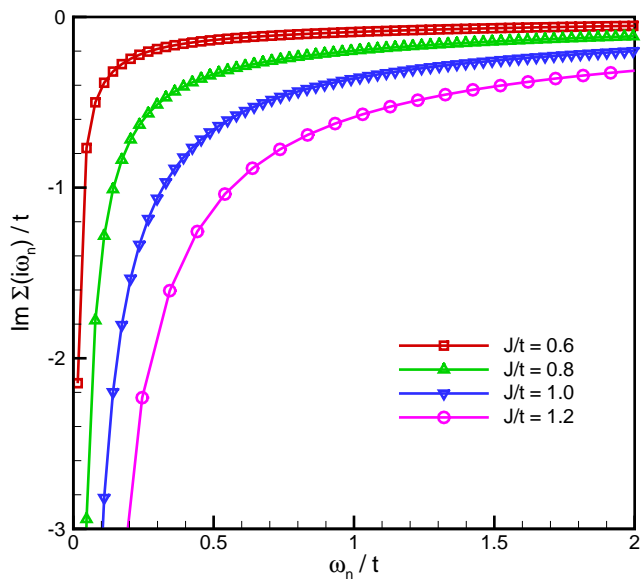}
\caption{(Color online) Imaginary part of the electron self-energy
\(\Sigma(i\omega_{n})\) for the half-filled single-channel Kondo
lattice model at \(T = 0.0, J/t = 0.6, 0.8, 1.0 \) and \(1.2\).}
\label{selfsingle}
\end{figure}

In this subsection, we consider the behavior of KLM in paramagnetic
phase. We first report results on SC-KLM. Our
calculations reproduces the results of Werner and Millis
\cite{Werner} obtained via stochastic QMC. In addition, the use of
ED allowed us to access lower temperatures and smaller frequencies.
Fig. \ref{selfsingle} shows the self-energies calculated for several
\(J\)-values at zero temperature.
\begin{figure}[htb]
\includegraphics[width=10cm,clip]{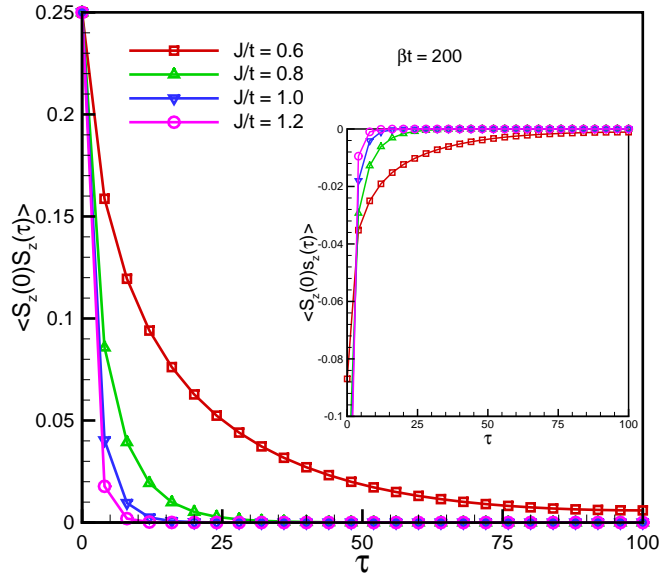}
\caption{(Color online) Imaginary time correlation function for the
local moments calculated at half filling for the \(J\)-values as
indicated and for \(\beta t = 200\). Inset shows conduction
spin-local spin correlation functions. Both spin-spin correlation
functions have exponentially decaying dependence on \(\tau\).  }
\label{correlation}
\end{figure}
As \(\omega_{n}\rightarrow 0\), even for smallest \(J\), the
imaginary part of the self-energies diverges. This shows the presence of a
charge gap at the half-filled SC-KLM and indicates that the system
is in the Kondo insulating phase. This is a quantum disordered phase
in which the conduction electrons are bound to local spins forming
spin singlets.
\begin{figure}[htb]
\includegraphics[width=10cm,clip]{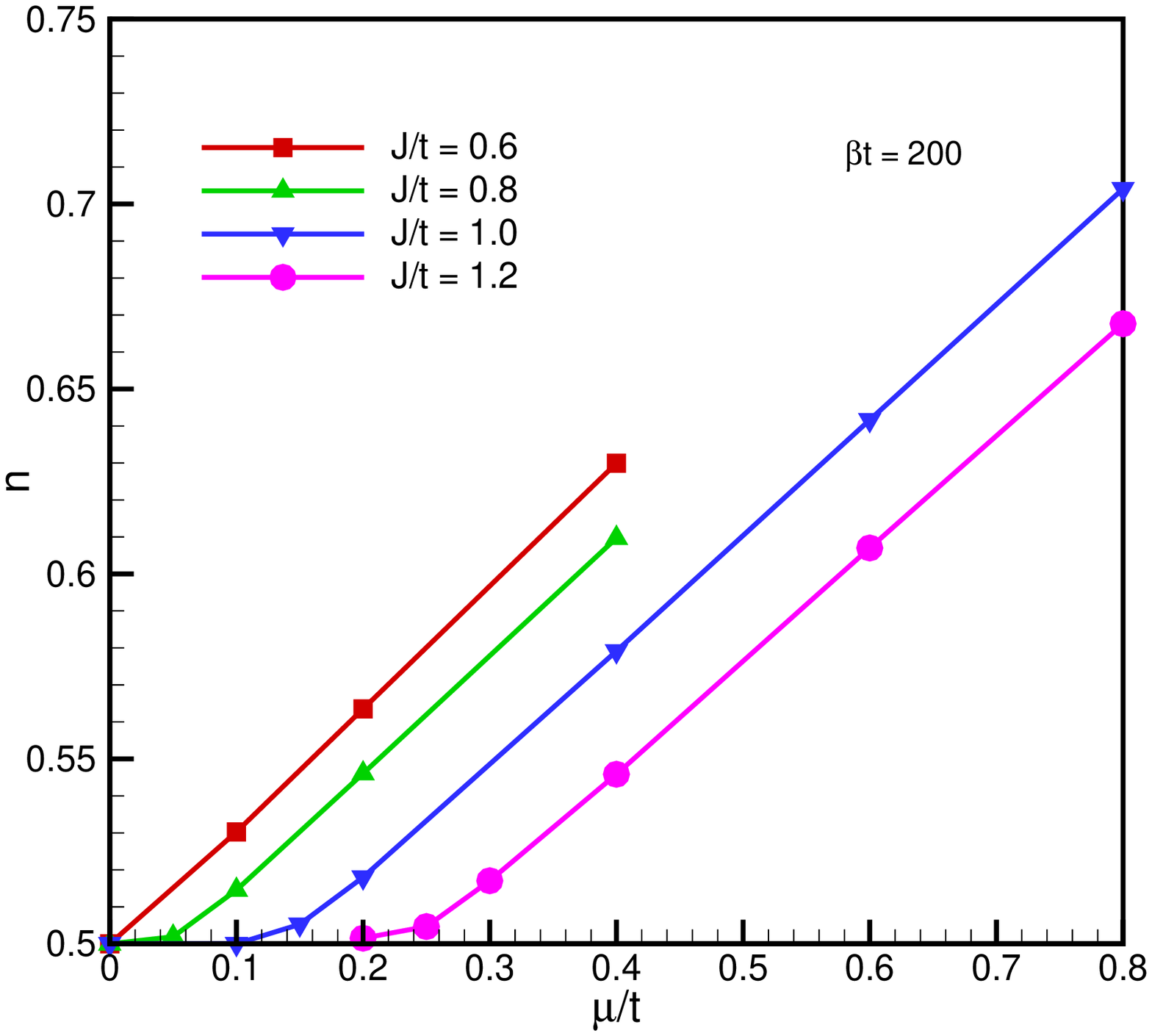}
\caption{(Color online) Density per spin plotted as a function of
chemical potential for \(\beta t = 200, J/t = 0.2, 0.6, 0.8\) and \(
1.0\). The data for \(J/t = 0.80\) and \(1.0\) are consistent with
the opening of a charge gap. } \label{density}
\end{figure}
The local spin-local spin correlation function \(\langle
S_{z}(0)S_{z}(\tau) \rangle\) and the local spin-conduction spin
correlation functions \(\langle S_{z}(0)s_{z}(\tau) \rangle \) at
\(\beta t = 200\) are shown in Fig. \ref{correlation}. The
correlations decay rapidly with time, consistent with the formation
of a gapped Kondo insulating state. Also the local spin-conduction
spin correlation (inset) indicates an antiparallel alignment
\(\langle S_{z}(0)s_{z}(\tau) \rangle < 0\). Fig. \ref{density}
shows the dependence of the particle number per spin, \(n\), on the
chemical potential, \(\mu\), for several \(J\)-values. This figure
is similar to Fig. 7 of Ref. \onlinecite{Werner}. However, because
of the use of ED in our calculations, the gap in the excitation
spectrum is quite evident for smaller values of \(J/t\).

\begin{figure}[htb]
\includegraphics[width=14cm,clip]{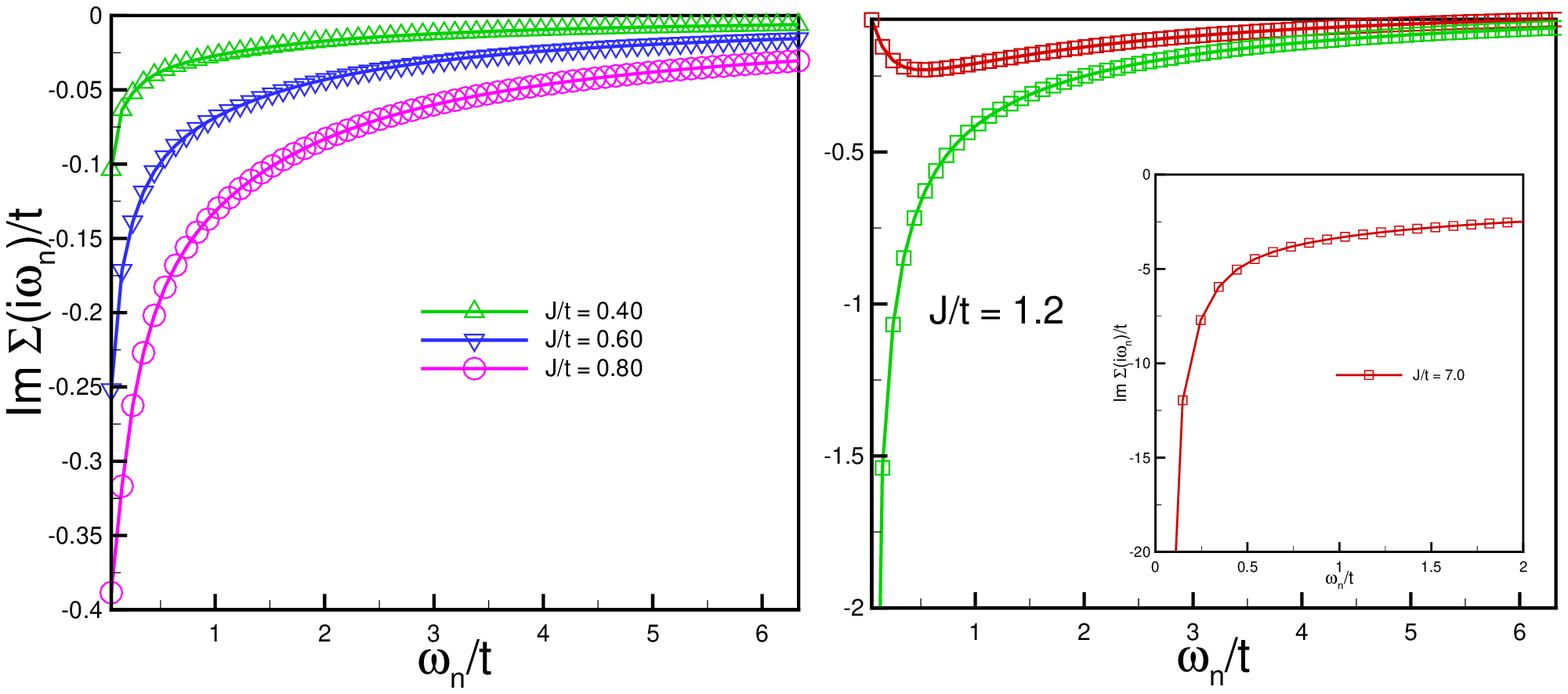}
\caption{(Color online) Imaginary part of the electron self-energy
\(\Sigma(i\omega_{n})\) for the half-filled TC-KLM at \(T = 0.0\)
and \( J/t = 0.2, 0.4, 0.6\) and \( 0.8\) (left panel) and \(T =
0.0, J/t = 1.2, 7.0\) (right panel). The spontaneous symmetry
breaking which takes place in \(J/t \geq 1\) is obvious from the
graph in the right panel. Further increase of the coupling strength
caused the electrons in both bands to form bound states with local
spins (see in the inset of right panel). } \label{selftwo}
\end{figure}

We next focus on TC-KLM. We have also calculated the imaginary part of
self-energies for the two-channel case at quarter-filling and found
that the situation in the two-channel model at quarter-filling is
different from single-channel case at half-filling. It is believed
that at the coupling strengths  \(J/t = 0.20, 0.60\) and \(1.20\),
electrons of different channels generate independent RKKY
interaction between the localized moments. Fig. \ref{selftwo} shows
the imaginary part of self-energies for TC-KLM at half-filling. In
the left panel of Fig. \ref{selftwo} our results for small coupling
strengths are shown, where the two bands show the same behavior.
These results are in qualitative agreement with the corresponding
results in Ref. \onlinecite{Jarrell}. The right panel shows the
result for larger coupling strengths. We see that with increasing
coupling strength, a symmetry breaking phase transition occurs as shown in
the right panel. As seen in the figure, due to the fact that the
self-energies assume two diffrenet values at a given Matsubara
frequency, we have obtained two diffrent bands where at low
frequencies one of them shows the metallic behavior and the other
one diverges indicating an insulating phase. The inset of the right
panel shows that at larger coupling strengths, conduction electrons
of both channels form bound states with local spins.

\subsection{Magnetic Ordering}
\begin{figure}[htb]
\includegraphics[width=14cm,clip]{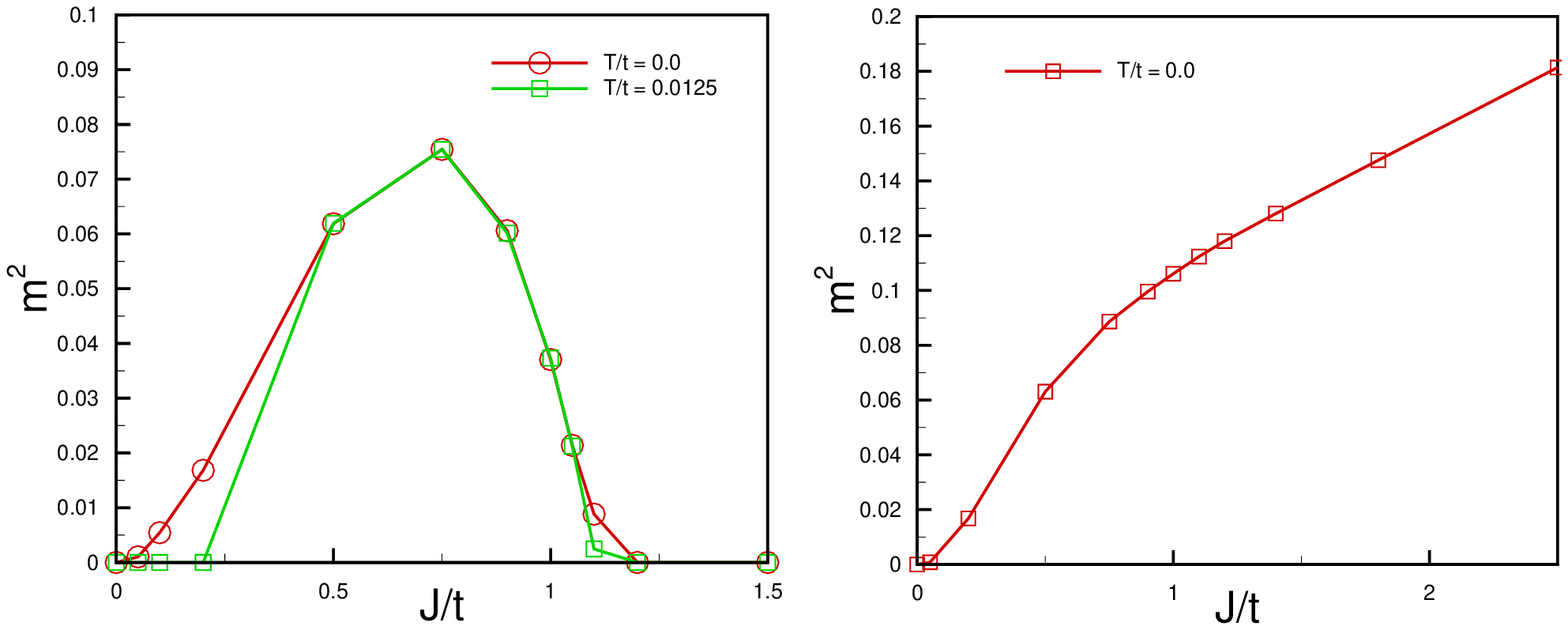}
\caption{(Color online) Staggered magnetization \(m = n_{\uparrow} -
n_{\downarrow}\) of the Kondo lattice model (half-filling, bipartite
lattice). Left panel: staggered magnetization of single-channel
Kondo lattice model as a function of \(J/t\) for \(T/t = 0.0,
0.0125\). There is an antiferromagnetic state at small coupling (for
sufficiently low temperature) and a quantum phase transition to a
paramagnetic insulator around \(J / t = 1.0\). Right panel:
staggered magnetization of two-channel Kondo lattice model as a
function of \(J/t\) for \(T/t = 0.0\). There is an antiferromagnetic
state for all physical coupling strengths. } \label{m}
\end{figure}
We now study the magnetic ordering phenomena characteristic of the
Kondo lattice. In the single channel KL we expect a quantum phase
transition to a singlet phase for \(J\) larger than a critical
value. In Fig. \ref{m} we show the staggered magnetization \(m =
n_{\uparrow} - n_{\downarrow}\) of the SC-KLM as a function of
\(J/t\) at \(T/t = 0.0, 0.0125\). On the small \(J\) side a strong
temperature dependence is evident, reflecting the strong \(J\)
dependence of the Neel temperature at weak coupling. As one can see,
by decreasing the temperature, the onset of magnetization shifts to
lower \(J\)-values. At \(J/t \geq 1\) the staggered magnetization
rapidly drops to zero for either of the two values of temperatures.
This is the quantum phase transition to the singlet Kondo insulator
phase. The right hand panel shows the staggered magnetization for
TC-KLM. This figure shows the existence of an antiferromagnetic
insulator for all physical coupling strengths.

\section{Concluding remarks}
We have studied the single- and two-channel KLM at quarter- and
half-filling using ED/DMFT approach. Compared to other frequently
used DMFT impurity solvers such as QMC, the ED has the advantage of
accessing very low temperatures and strong couplings. Our results
for the single-channel case showed the presence of a charge gap at
the half-filling and indicates that the system is in the Kondo
insulating phase. In the two-channel case we find a symmetry
breaking phase transition with increasing coupling strength.

As mentioned in the introduction, the nature of the ground state of
the KLM results from the interplay between the magnetic RKKY
interaction among the localized spins and the Kondo effect screening
of these spins. In fact, we would expect to observe a strong
competition between the RKKY interaction and the Kondo screening
effect whenever the N\'{e}el and Kondo temperatures are rather close
to each other.

Recent experiments in \(\mathrm{Ce}\mathrm{In}_{3}\) and
\(\mathrm{CePd}_{2}\mathrm{Si}_{2}\) exhibit such a strong
competition as evidenced by their nearly the same N\'{e}el and Kondo
temperatures. Moreover, these two heavy fermion compounds show, as we
expect, an antiferromagnetic long range order \cite{Acquarone} (See
the Refs. therein). In order to investigate the properties of these
systems, one might supplement the KLM with a Heisenberg hamiltonian
for the localized spins, i.e., by including the antiferromagnetic
exchange, \(J_{AF}\), between core spins. In the future work, we plan
to use such a modified model employing the more sophisticated
cluster dynamical mean field theory.

\begin{acknowledgments}
We wish to acknowledge useful discussions with R. Asgari.
\end{acknowledgments}


\begin{thebibliography}{99}
\bibitem{ED} We solve the effective impurity problem by truncating
the sum on bath levels to a small number of terms \(n_{s}\), so that
the Hilbert space is small enough to use, e.g., the Lanczos
algorithm to compute the \(T=0\) Green's function. In all our
calculations presented here for SC-KLM the convergence of turncation
has been checked, and in the case of TC-KLM we have used four bath
levels per impurity level (\(n_{s} = 8\)). To check the accuracy of
this approximation we have repeated the evalution of our results
with five bath levels per impurity level (\(n_{s} = 10\)) and found
fairly good agrement with \(n_{s} = 8\).
\end{thebibliography}
\end{document}